# Broadband Enhancement of On-chip Single Photon Extraction via Tilted Hyperbolic Metamaterials

Lian Shen[1,2,#], Xiao Lin[3,#,*], Mikhail Shalaginov[4], Tony Low[5], Xianmin Zhang[1,2], Baile Zhang[3,6,*], and Hongsheng Chen[1,2,*]

[1]Interdisciplinary Center for Quantum Information, State Key Laboratory of Modern Optical Instrumentation, College of Information Science and Electronic Engineering, Zhejiang University, Hangzhou 310027, China.

[2]Key Lab. of Advanced Micro/Nano Electronic Devices & Smart Systems of Zhejiang, The Electromagnetics Academy at Zhejiang University, Zhejiang University, Hangzhou 310027, China.

[3]Division of Physics and Applied Physics, School of Physical and Mathematical Sciences, Nanyang Technological University, Singapore 637371, Singapore

[4]Department of Materials Science and Engineering, Massachusetts Institute of Technology, 77 Massachusetts Avenue, Cambridge, MA 02139, USA

[5]Department of Electrical and Computer Engineering, University of Minnesota, Minneapolis, Minnesota 55455, USA

[6]Centre for Disruptive Photonic Technologies (CDPT), School of Physical and Mathematical Sciences, Nanyang Technological University, Singapore 637371, Singapore

[#]These authors contributed equally to this work.

[*]Corresponding author. E-mail: xiaolinbnwj@ntu.edu.sg; blzhang@ntu.edu.sg; hansomchen@zju.edu.cn

**A fundamental building block for on-chip quantum photonics is a single-photon source with high repetition rates, which can enable many applications such as high-speed quantum communication and quantum information processing. Ideally, such single photon sources would then require a large on-chip photon extraction decay rate, namely the rate of excited photons coupled into nanofibers or waveguides, over a broad spectral range. However, this goal has remained elusive till date. Here we propose a feasible scheme to enhance the on-chip photon extraction decay rate of quantum emitters, through the tilting of the optical axis of hyperbolic metamaterials with respect to the end-facet of nanofibers. Importantly, the revealed scheme is applicable to arbitrarily orientated quantum emitters over a broad spectral range, e.g., up to ~80 nm for visible light. The underlying physics relies on the emerging unique feature of hyperbolic metamaterials if their optical axis is judiciously tilted. That is, their supported high-$k$ (i.e., wavevector) hyperbolic eigenmodes, which are intrinsically confined inside them if their optical axis is un-tilted, can now become momentum-matched with the guided modes of nanofibers, and more importantly, they can safely couple into nanofibers almost *without* reflection.**

The on-chip source of single photons is of extreme importance for the exploration and development of integrated quantum information technology, including quantum cryptography [1], quantum computation [2,3] and quantum communication [4-6], since existing quantum networks transmit information through single photons. One of the key characteristics of on-chip single photon sources is the repetition rate. This is because a higher repetition rate allows higher bit rates in quantum communication and faster readout of stationary qubits [7-10]. To be specific, the repetition rate is determined by the photon extraction decay rate of quantum emitters into nanofibers, which is equal to the product of emitter's spontaneous emission rate and the coupling efficiency of excited photons into nanofibers or waveguides. As such, it is highly desirable to enhance the on-chip photon extraction decay rate for single photon sources.

Although there are extensive studies about enhancing the spontaneous emission rate and the coupling efficiency [11-27], only a few proposals have been put forward to enhance the on-chip photon extraction decay rate [20-33]. These proposals either have a weak enhancement of the photon extraction decay rate or can work only at a narrow frequency range [20-33]. Correspondingly, the realization of large on-chip photon extraction decay rates over a broad spectral range still remains an open challenge. To clearly illustrate this issue, those proposals based on hyperbolic metamaterials are briefly analyzed as a typical example below. For hyperbolic metamaterials, placing a quantum emitter in their vicinity generally provides a drastic enhancement of the spontaneous emission rate over a broad spectral range, due to their broadband large photonic density-of-states [34-39]. However, their large photonic density-of-states cannot directly lead to a high photon extraction decay rate, due to the low coupling efficiency. To be specific, the coupling efficiency is severely limited by the fact that the supported high-*k* (or wavevector) eigenmodes generally cannot propagate to the far field but are intrinsically confined inside the hyperbolic metamaterials [37-39]. As such, the hyperbolic metamaterials, including those that are nano-patterned or adiabatically tapered [28-31], are thought to be less practical to achieve the broadband enhancement of on-chip photon extraction decay rates.

Here we introduce a feasible scheme where the above intrinsic limitations in hyperbolic metamaterials can be overcome, thus enabling a flexible enhancement of the on-chip photon extraction decay rate over a broad spectral range. To be specific, such a capability is enabled by the judicious tilting of the optical axis of hyperbolic metamaterials with respect to the end-facet of nanofibers; see the structural schematic in Fig. 1(a). Due to the tilted optical axis, the high-*k* eigenmodes in the hyperbolic metamaterials can become momentum-matched with the guided modes in nanofibers, and more importantly, their reflection would be suppressed at the interface between hyperbolic metamaterials and nanofibers. In other words, the high-*k* eigenmodes can now safely couple into nanofibers. This way, the tilted hyperbolic metamaterials here are distinctly different from un-tilted hyperbolic metamaterials [28-31], and they become practical for the broadband enhancement of on-chip single photon extraction. Another advantage is that the proposed scheme is applicable for quantum emitters with arbitrary orientation. Our work thus represents a vital step towards the implementation of spectrally broad single photon sources with high repetition rates for on-chip quantum networks.

We start with the analysis of the corresponding underlying mechanism, from the perspective of the isofrequency contour of eigenmodes in hyperbolic metamaterials. For the hyperbolic metamaterial, its optical axis has an angle $\theta$ with respect to the normal vector $\hat{n}$ ($\hat{n}||\hat{z}$) of the end-facet (parallel to the $x$-$y$ plane) of nanofibers [Fig. 1(a)], and it has a relative permittivity of $[\varepsilon_{||}, \varepsilon_{||}, \varepsilon_{\perp}]$, where $\varepsilon_{||}$ and $\varepsilon_{\perp}$ are the components parallel and perpendicular to the optical axis, respectively. This way, the isofrequency contour in the $x$-$z$ plane can be described by

$$(\varepsilon_{||} \sin^2\theta + \varepsilon_{\perp} \cos^2\theta)k_z^2 + 2(\varepsilon_{||} - \varepsilon_{\perp})\sin\theta \cos\theta\, k_x k_z + (\varepsilon_{||}\cos^2\theta + \varepsilon_{\perp}\sin^2\theta)k_x^2 = k_0^2 \varepsilon_{||}\varepsilon_{\perp} \quad (1)$$

where $k_x$ and $k_z$ are the components of wavevector $\bar{k}$ parallel and perpendicular to the end-facet of nanofibers [Fig. 1(a)], respectively; $k = |\bar{k}| = \sqrt{k_x^2 + k_z^2}$; $k_0 = \omega/c$; $c$ is the light speed in free space. Below we set $\text{Re}(\varepsilon_{||}) > 0$ and $\text{Re}(\varepsilon_{\perp}) < 0$.

We emphasized that the term related to $k_z^2$ in equation (1) disappears, if $\varepsilon_{||}\sin^2\theta + \varepsilon_{\perp}\cos^2\theta = 0$ or $\theta = \varphi_c$, where $\varphi_c = \tan^{-1}(\sqrt{-\varepsilon_{\perp}/\varepsilon_{||}})$ is the critical angle between the optical axis and the asymptotic line of hyperbolic isofrequency contours [Fig. 1(b)]. As a result, there is always only unique solution of $k_z$ for an arbitrary value of $k_x$ in equation (1) if $\theta = \varphi_c$ [Fig. 1(c)]. In contrast, there is always two solutions of $k_z$ for an arbitrary value of $k_x$ if $\theta = 0$ [Fig. 1(b)].

The unique feature of hyperbolic isofrequency contour in Fig. 1(c) can lead to two prominent ways to enhance the on-chip photon extraction decay rate in Fig. 1(a). First, the supported high-$k$ (i.e., $k \gg k_0$) hyperbolic modes become to have a small value of $k_x$ if $\theta = \varphi_c$. To be specific, we have $|\text{Re}(k_x/k_0)| \leq n_{\text{fiber}}$ even if $k \to \infty$ in the colored region in Fig. 1(c), where $n_{\text{fiber}}$ is the refractive index of the constituent material for the nanofiber (e.g., if Si$_3$N$_4$ is adopted in Fig. 1(a), $n_{\text{fiber}} = n_{\text{Si}_3\text{N}_4}$). Then from the phase matching condition at the end-facet of nanofibers, these high-$k$ hyperbolic modes now have the chance to couple into the guided modes of the nanofiber. In contrast, most high-$k$ hyperbolic modes in Fig. 1(b) have $|\text{Re}(k_x/k_0)| > n_{\text{fiber}}$. They are thus intrinsically confined within the un-tilted metamaterials and cannot couple into the nanofiber.

Second, the reflection of hyperbolic modes, including those high-$k$ modes, at the end-facet of the nanofiber will be completely suppressed. This is because for a determined $k_x$ in Fig. 1(c), there is only one hyperbolic mode, which corresponds to the incident hyperbolic mode, supported by the metamaterial. In other words, there will be no reflected propagating hyperbolic mode, since the reflected fields are evanescent with respect to the interface. This unavoidably leads to a high transformation of arbitrary incident hyperbolic modes in the shaded region of Fig. 1(c) into the guided modes of nanofibers. In contrast, for the un-tilted hyperbolic metamaterials in Fig. 1(b), due to the existence of reflected propagating hyperbolic modes, the reflection of an incident hyperbolic mode (including the low-$k$ ones) cannot be avoided and is significant for high-$k$ modes.

From the above analyses, it is then straightforward to use the emerging unique feature of tilted hyperbolic metamaterials in Fig. 1(c) to enhance the on-chip photon extraction decay rate. This can be done, for example, simply by positioning a quantum emitter very close to a thin slab of the hyperbolic metamaterial, which is integrated with a nanofiber [Fig. 1(a)]. Here we focus on the weak coupling regime, and the quantum emitter is modeled by a dipole source [40]; see supporting information. Since the hyperbolic metamaterial has a broadband extremely-large photonic density-of-states [34-39], the emitter has a large spontaneous emission rate $\gamma_{\text{total}}$ or a large Purcell enhancement $\gamma_{\text{total}}/\gamma_0$ (e.g., $> 10^5$ in Fig. S2), where $\gamma_0$ is the spontaneous emission rate of the quantum emitter in free space.

If the material loss is neglected, the on-chip photon extraction decay rate $\gamma_{\text{fiber}}$ at a specific wavelength in Fig. 1(a) can be extremely large, and in principle, it can reach a value having the same order of magnitude as $\gamma_{\text{total}}$ (i.e., $\gamma_{\text{fiber}} \sim \gamma_{\text{total}}$). This is enabled by the revealed capability of tilted hyperbolic metamaterials to efficiently transform the excited hyperbolic modes in the shaded region of Fig. 1(c) into the guided modes of the nanofibers. It is then reasonable to expect $\gamma_{\text{fiber}}$ to be quite large, e.g., $\gamma_{\text{fiber}}/\gamma_0 > 10^5$, in a broad spectral range, if the tilted hyperbolic metamaterial is transparent.

However, the material loss is unavoidable in practical passive hyperbolic metamaterials. We note that the realistic material loss can lead to a large reduction of $\gamma_{\text{fiber}}$, since the energy of excited high-$k$ modes will be largely degraded before they arrive at the end-facet of nanofibers. We then proceed to the study of $\gamma_{\text{fiber}}$ in Figs. 2-3, with the consideration of realistic material losses in tilted hyperbolic metamaterials. Figures 2-3 show that we can still achieve $\gamma_{\text{fiber}} \gg \gamma_0$ (e.g., $\gamma_{\text{fiber}} > 100\gamma_0$) over a broad spectral range, although we generally have $\gamma_{\text{fiber}} \ll \gamma_{\text{total}}$, instead of $\gamma_{\text{fiber}} \sim \gamma_{\text{total}}$, for realistic cases [Fig. S2].

We begin our numerical study of $\gamma_{\text{fiber}}$ at a specific wavelength in Fig. 2; see the computation detail of $\gamma_{\text{fiber}}$ in supporting information. For hyperbolic metamaterials, they are effectively constructed through alternating layers of metals and dielectrics. As a conceptual demonstration, the experimental data of permittivity of silver and silica [41] are adopted, and the metallic filling fraction is 50%; see the detailed strategy of structural design in supporting information. Due to the dispersion of constituent materials (e.g., metal), the critical angle $\varphi_c(\lambda)$ of the designed hyperbolic metamaterial is sensitive to the wavelength [Fig. 2(a)].

Figure 2(b) shows $\gamma_{\text{fiber}}$ as a function of the tilted angle $\theta$. The working wavelength (in free space) of $\lambda_1 = 685$ nm is chosen for illustration, since it is within the radiation spectrum in which many of the currently considered quantum emitters operate, such as the range of 575 to 785 nm for the nitrogen-vacancy (NV) center in nano-diamonds [42]. The maximal $\gamma_{\text{fiber}} \approx 460\gamma_0$ appears for the quantum emitter oriented along the $z$ direction if $\theta = \varphi_c(\lambda_1)$ in Fig. 2(b), where we have $\varphi_c(\lambda_1) = 36°$ from Fig. 2(a). Correspondingly, if $\theta = \varphi_c(\lambda_1)$, we find significant radiation fields being extracted into the nanofiber at $\lambda_1$ in Fig. 2(c). As complementary information, $\gamma_{\text{fiber}}$ is also studied for the quantum emitter oriented along the $x$ or $y$ direction in Fig. S2. Importantly, the large value of $\gamma_{\text{fiber}} > 400\gamma_0$ is always achievable if $\theta = \varphi_c(\lambda_1)$, independent of the orientation of quantum emitters. From Fig. 2 and Fig. S2, it is reasonable to argue that the proposed strategy to enhance $\gamma_{\text{fiber}}$ via tilted (transparent) hyperbolic metamaterials in Fig. 1 works well for scenarios with reasonable amount of loss.

Moreover, the proposed strategy in Fig. 1 is also applicable for the realistic lossy case over a broad spectral range, as shown in Fig. 3. To illustrate this point, Fig. 3 shows $\gamma_{\text{fiber}}$ as a function of the wavelength ($\lambda \in [575, 785]$ nm) and the tilted angle $\theta$. If $\theta = \varphi_c(\lambda_1)$, $\gamma_{\text{fiber}} > 100\gamma_0$ can be achieved over a continuous wavelength range with a bandwidth of ~80 nm [Fig. 3]. Note that the relative bandwidth, namely the bandwidth normalized by the central working wavelength, reaches $12\% = \frac{80\ nm}{\lambda_1}$.

On the other hand, it is noted that the feature of $\gamma_{\text{fiber}}$ as a function of wavelength at $\theta = \varphi_c(\lambda_1) \pm 1°$ is similar to that at $\theta = \varphi_c(\lambda_1)$ in Fig. 3. Such a phenomenon indicates that due to materials losses, it is the excited hyperbolic mode with a finite large $k$ ($\leq k_{\text{max}}$), instead of the one with extremely large $k$ ($> k_{max}$), that makes the main contribution to the enhancement of $\gamma_{fiber}$. This way, the critical requirement of $\theta = \varphi_c(\lambda_1)$ in Fig. 1(c), which is applied to achieve large $\gamma_{fiber}$ over a broad spectral range around $\lambda_1$, is not particularly stringent for the realistic lossy case in Fig. 3. In other words, there is a certain tolerance on the choice of the tilted angle for the lossy cases. Such a loose requirement on the tilted angle in Fig. 3 may facilitate the practical implementation of the proposed structure in Fig. 1.

Last but not least, the permittivity of hyperbolic metamaterials can be described by the effective medium theory in an approximate way [43-45] or by the Bloch theorem in an accurate way [46]. Since $\gamma_{fiber}$ for the realistic lossy cases is mainly related to the excited hyperbolic modes with $k \leq k_{max}$, it is reasonable and convenient to apply the effective medium theory in Figs. 2-3; see more discussion in Fig. S3.

In summary, we have proposed a simple yet universal scheme to enhance $\gamma_{fiber}$ in a broad spectral range, e.g., a bandwidth of ~80 nm in the visible regime, via tilted hyperbolic metamaterials. For tilted hyperbolic metamaterials, we let the tilted angle $\theta$ of their optical axis to be equal to their critical angle $\varphi_c$. Such a judicious geometrical tilting gives rise to some unique features of the hyperbolic isofrequency contour; in particular, it allows the excited high-$k$ hyperbolic modes to efficiently couple into the guided modes of the nanofiber almost without reflection. Two further advantages of the proposed scheme are the loose requirement on the tilted angle and the orientation of quantum emitters, which may facilitate its practical implementation. Our work also triggers many interesting open questions. As a prototypical example, if a quantum emitter is positioned close to a realistic hyperbolic metamaterial, the possibility to realize $\gamma_{\text{fiber}}$ with its value in the same order of magnitude as $\gamma_{\text{total}}$, namely the simultaneous realization of a extremely-large spontaneous emission rate and a high coupling efficiency via hyperbolic metamaterials, still remains elusive.

**Acknowledgements**
The work was sponsored by the National Natural Science Foundation of China (NNSFC) under Grants No. 61625502, No. 61574127, No. 61975176, No. 61905216, China Postdoctoral Science Foundation (2018M632462), Nanyang Technological University for NAP Start-Up Grant and the Singapore Ministry of Education (Grant No. MOE2018-T2-1-022 (S), MOE2016-T3-1-006 and Tier 1 RG174/16 (S)).

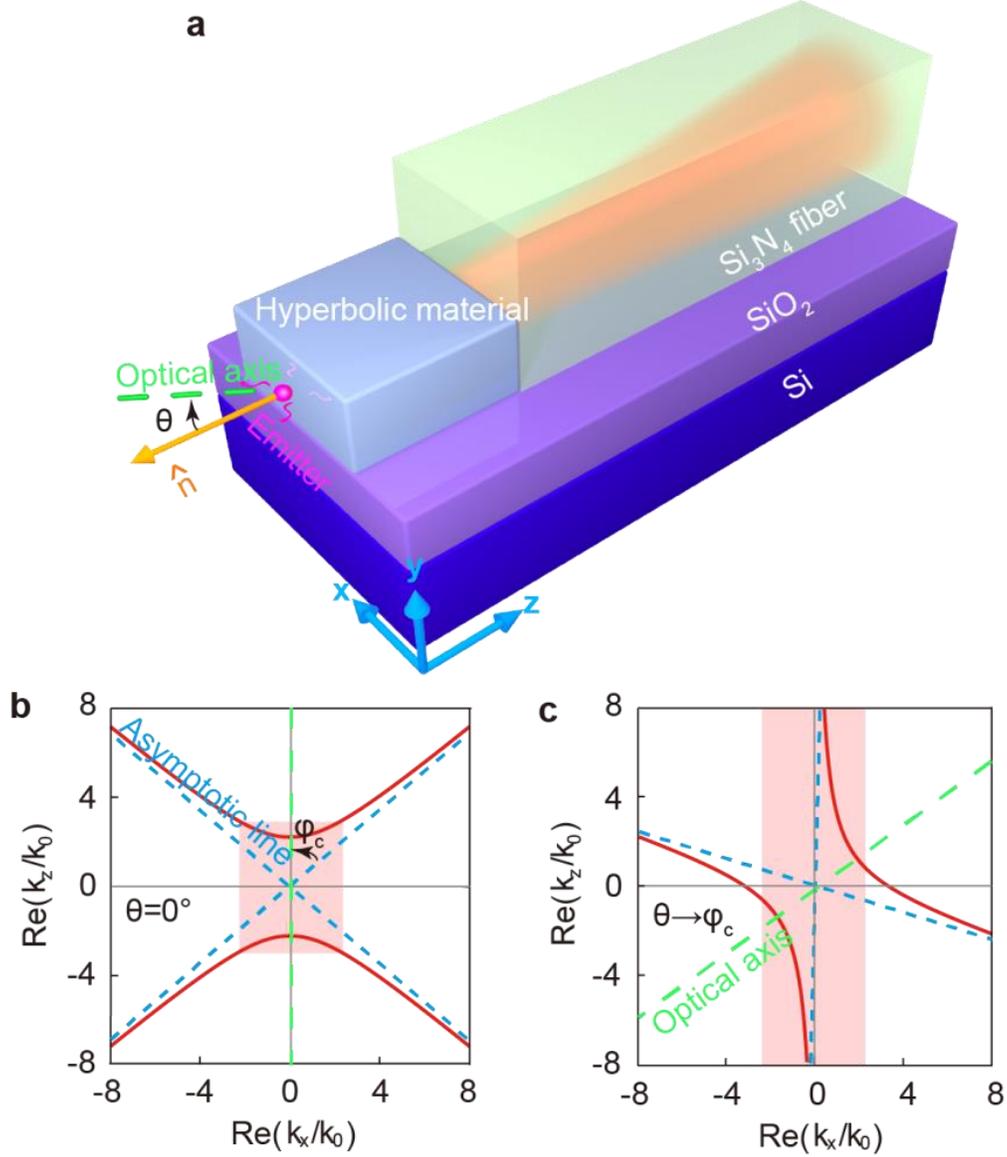

**Figure 1. Schematic of enhancing the on-chip single photon extraction decay rate via tilted hyperbolic metamaterials.** (**a**) Structural setup. A quantum emitter, modeled by a dipole, is positioned very close to a hyperbolic metamaterial, which is integrated with a nanofiber and whose optical axis is tilted by an angle of $\theta$ with respect to the normal vector $\hat{n} = -\hat{z}$ of the interface. (**b,c**) Isofrequency contour of transparent hyperbolic metamaterials with (b) $\theta = 0$ and (c) $\theta \to \varphi_c$, where $\varphi_c$ is the angle between the optical axis and the asymptotic line of hyperbolic isofrequency contours. The colored or shaded region in (b,c) has $|\text{Re}(k_x/k_0)| \leq n_{Si_3N_4}$; the excited hyperbolic modes inside this region in (c) are able to efficiently couple into the nanofiber almost without reflection.

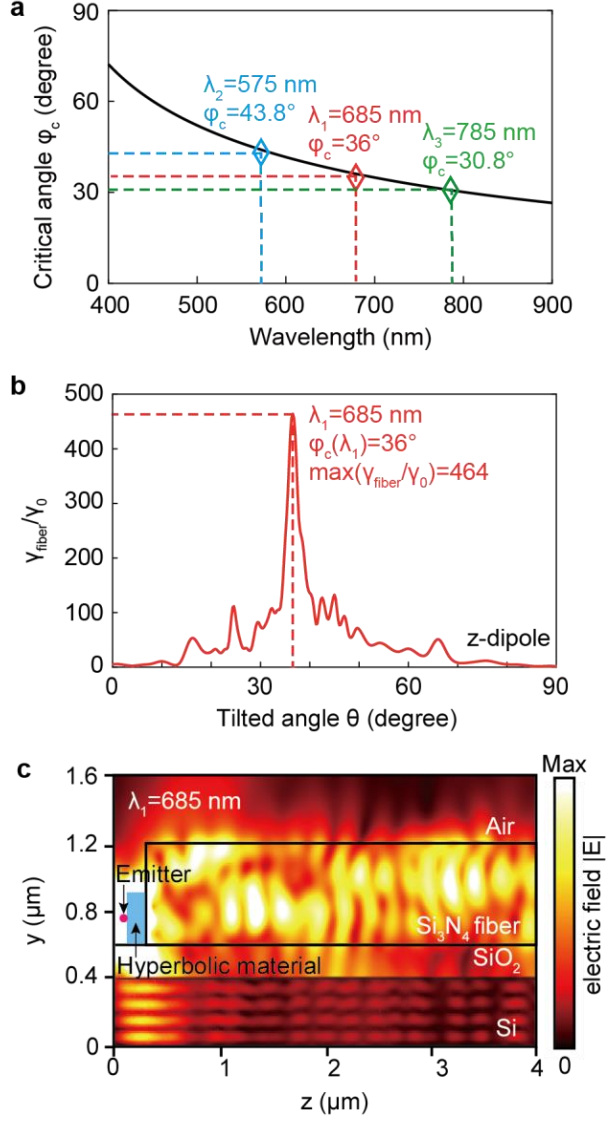

**Figure 2. Numerical demonstration of the enhancement of on-chip single photon extraction via tilted hyperbolic metamaterials with realistic material losses.** (a) Dependence of $\varphi_c$ on the wavelength. (b) Normalized photon extraction rate $\gamma_{\text{fiber}}/\gamma_0$ of a single emitter into nanofibers as a function of the tilted angle $\theta$. $\gamma_0$ is the spontaneous emission rate of a quantum emitter in free space. (c) Excited field distribution of a dipole-like emitter oriented along $z$ direction, where $\theta = \varphi_c(\lambda_1) = 36^o$. The metamaterial and nanofiber have a cross-section of 600×300 nm² and 600×600 nm² in the $xy$ plane, respectively. In the $z$ direction, the metamaterial and SiO$_2$ film have a thickness of 200 nm and 100 nm, respectively. $\varepsilon_{Si_3N_4} = n^2_{\varepsilon_{Si_3N_4}} = 4.37$ at 685 nm. The other structural setup is the same as Fig. 1c.

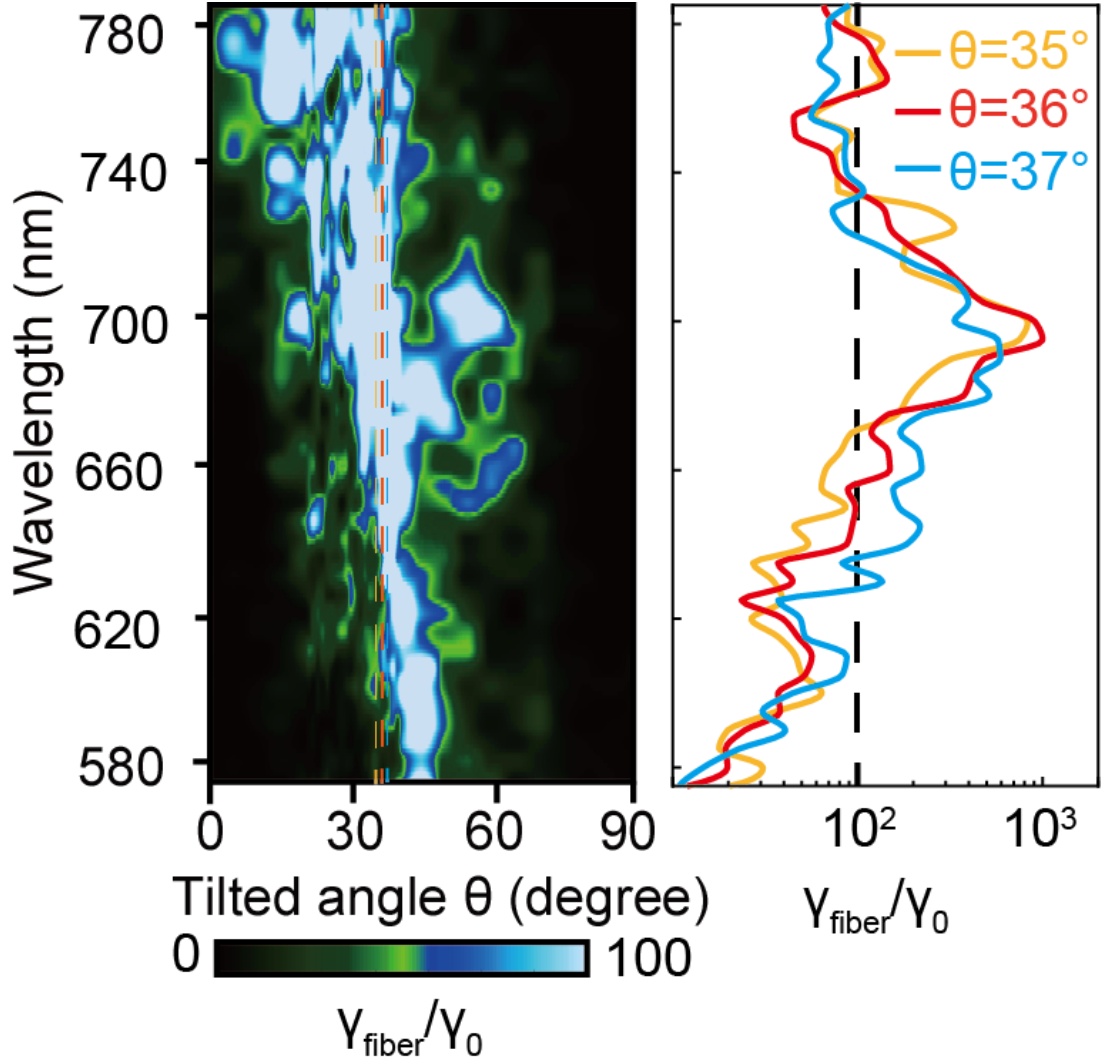

**Figure 3. Broadband enhancement of the on-chip single photon extraction via tilted hyperbolic metamaterials.** The left panel shows the normalized on-chip photon extraction decay rates $\gamma_{\text{fiber}}/\gamma_0$ as a function of wavelength and the tilted angle $\theta$. In particular, $\gamma_{\text{fiber}}/\gamma_0$ at $\theta = \varphi_c(\lambda_1)$ or $\varphi_c(\lambda_1) \pm 1°$ are highlighted as a function of wavelength in the right panel, where $\varphi_c(\lambda_1) = 36°$. The bandwidth for $\gamma_{\text{fiber}}/\gamma_0 \geq 100$ can reach ~80 nm for visible light if $\theta \to \varphi_c(\lambda_1)$. The basic setup is the same as Fig. 1c.